\documentclass[twocolumn,showpacs,preprintnumbers,amsmath,amssymb,prb]{revtex4}
\usepackage{graphicx}
\usepackage{dcolumn}
\usepackage{bm}
\usepackage{longtable}
\usepackage{amsmath}
\usepackage{units}
\usepackage{todonotes}
\usepackage{color} 
\usepackage{braket}

\usepackage{hyperref}
\hypersetup{colorlinks=true, linkcolor=blue, citecolor=red, urlcolor=magenta, pdfauthor={M. Fechner}}

\renewcommand{\vec}[1]{\bm{#1}}

\newcommand{\PH}[2]{$#1_{\textnormal{#2}}$}
\newcommand{\AMP}[2]{Q_{\mathrm{#1}_\mathrm{#2}}}
\newcommand{\freq}[2]{\omega_{\textnormal{#1}_{#2}}}
\newcommand{\YBCO}{YBa$_2$Cu$_3$O$_7$}

\newcommand{\YBCOsixfive}{YBa$_2$Cu$_3$O$_{6.5+x}$}

\begin{document}
\title{Effects of intense optical phonon pumping on the structure and electronic properties of yttrium barium copper oxide}
\author{M. Fechner}
\email{michael.fechner@mat.ethz.ch}
\affiliation{Materials Theory, ETH Zurich, CH-8093 Z\"urich, Switzerland }
\author{N. A. Spaldin}
\affiliation{Materials Theory, ETH Zurich, CH-8093 Z\"urich, Switzerland }

\begin{abstract}
We investigate the structural modulations induced by optical excitation of a polar phonon mode in \YBCO, using first-principles calculations based on density functional theory. We focus on the intense-excitation regime in which we expect that fourth-order phonon-phonon coupling terms dominate, and model the structural modulations induced by pulses of such intensity. Our calculations of the phonon-phonon anharmonicities confirm that the cubic coupling between modes, shown in earlier work to cause a quasi-static change in the apical O - Cu distance and a buckling of the CuO$_2$ planes, is the leading contribution at moderate pump strengths. At higher pump strengths ($\sim$\unit[10]{MV/cm}) the previously neglected quartic couplings become relevant and produce an additional shearing of the CuO$_2$ planes. Finally, we analyze the changes in the electronic and magnetic properties associated with the induced structural changes. 
\end{abstract}

\maketitle

\section{INTRODUCTION}
Ultrafast modulation of crystal structures using THz radiation is an emerging technique in condensed matter physics to study the interplay of structural and electronic properties \cite{Orenstein:2012bz}. For example, control of electronic phases has been achieved in perovskite-structure manganites by selective pumping of phonon modes \cite{Forst:2011ep,Rini:2007hc}, and driving of spin dynamics has been demonstrated through excitation of coupled spin-phonon modes \cite{Kubacka:2014bf}. Particularly intriguing was the recent report of induced coherent transport -- a possible signature of superconductivity -- far above the usual superconducting $T_c$ in underdoped YBaCu$_3$O$_{6+\delta}$ on optical pumping of the IR-active \PH{B}{1u} mode at \unit[670]{cm$^{-1}} (\unit[20]{THz})$ \cite{Kaiser:2012ux}. The eigenvector of this mode consists of the in-phase displacement of the apical oxygen atoms of the Cu-O planes along the $c$ axis (Fig.~\ref{fig_modes} (a)) with an associated change of the Cu - apical oxygen distance, a parameter that has often been suggested to correlate with superconductivity \cite{Pickett:1989tk}. The enhanced coherence was therefore associated with an increase (decrease) in inter- (intra-) bilayer Josephson tunneling strength \cite{Hu:2013vw}. 

An important breakthrough in understanding this observation followed with the determination, through a combined ultra-fast x-ray diffraction (XRD) and \textit{ab-initio} density functional theory (DFT) study, of the transient crystal structure during the optical pumping process \cite{Mankowsky:2014vt}. As expected, oscillating staggered dilations/contractions of the Cu-O intra- (inter-)bilayer distances corresponding to the displacements of the \PH{B}{1u} mode were observed. In addition, anisotropic changes in the buckling of the in-plane Cu-O bonds were found, and shown to result from a cubic coupling of the \PH{A}{g} symmetry mode corresponding to this additional distortion to the square of the pumped \PH{B}{1u} mode \cite{Subedi:2013uh}.

The basic physics of THz-radiation-induced structural distortion, often referred to as non-linear phononics \cite{Subedi:2013uh,Forst:2011ep}, can be seen from analyzing the phonon-phonon interactions. For low field strengths, excitation of a phonon results only in harmonic oscillation of the atoms around their equilibrium positions, described by the harmonic Hamiltonian
\begin{equation}
H^{\mathrm{har}} = \frac{\omega^2 \AMP{}{}^2} {2} \quad .
\end{equation}
The atomic displacements are determined by the normal coordinate of the mode, $\AMP{}{}$, which in turn is the eigenvector of the dynamical matrix, and the corresponding eigenvalue, $\omega$, gives the mode frequency. Phonons of different energy are orthogonal and do not interact in the harmonic approximation, and importantly there is no change in the time-averaged structure with phonon excitation ($\Braket{Q}_T=0$). 

At higher field strengths and larger amplitudes, however, anharmonic phonon-phonon interactions become noticeable. Using the notation of Ref.~\onlinecite{Mankowsky:2014vt}, the anharmonic terms in the Hamiltonian can be written as 
\begin{equation}\label{eq_anharmonic}
H^{\mathrm{anh}} = -a_{3}  \AMP{IR}{}^2  \AMP{G}{}  -a_{4} \AMP{IR}{}^2  \AMP{G}{}^2 + ...
\end{equation}
considering terms only up to quartic order. Here $\AMP{IR}{}$ labels an infra-red active mode that can be excited by an optical phonon, and $\AMP{G}{}$ is a general polar or non-polar mode. We note that the term containing $a_{3}$ is only allowed if the symmetry of $\AMP{G}{}$ is given by an $A_1$ irreducible representation of the crystal structure point group.

In the case of the high-$T_c$ superconducting cuprates, which are centrosymmetric, the third-order coupling is sizable for several modes $\AMP{G}{}$ of \PH{A}{g} symmetry. As mentioned above, this cubic coupling to \PH{A}{g} modes was shown to be responsible for the shift of the mean atomic displacements measured using femtosecond x-ray diffraction in Ref.~\onlinecite{Mankowsky:2014vt}. Note that, in the experiments of Ref.~\onlinecite{Mankowsky:2014vt}, the system was excited using mid-infrared optical pulses of $\sim$\unit[300]{fs} duration with a maximum fluence of \unit[4]{mJ/cm$^2$}, which corresponds to a peak electric field of $\sim$\unit[3]{MV/cm}. This field strength suggests a peak amplitude of the \PH{B}{1u} mode corresponding to a \unit[2.2]{pm} increase in the apical oxygen-Cu distance. 

The product of $\AMP{IR}{}^2  \AMP{G}{}^2$ is totally symmetric and hence fourth-order coupling of any IR pumped mode to all other phonons is always allowed by symmetry. As mentioned in Ref.~\onlinecite{Mankowsky:2015gv}, analysis of the equation of motion shows that, for small amplitude excitations, the effect of this coupling is renormalization of the frequency of the $\AMP{G}{}$ mode by
\begin{equation}\label{eq_mode_renormalization}
\omega_\mathrm{G}' = \omega_\mathrm{G} \sqrt{1-\frac{2 a_2}{\omega^2_\mathrm{G}}\AMP{IR}{}^2} \quad .
\end{equation}
Eqn.~\eqref{eq_mode_renormalization} also reveals that this quartic interaction can induce a softening of the $\AMP{G}{}$ mode and an associated lattice instability, in the case of large $\AMP{IR}{}$ amplitude, strong coupling and low frequency of the mode $\AMP{G}{}$. For the excitation strengths of Ref.~\onlinecite{Mankowsky:2015gv}, fourth-order coupling effects were not observed.

The aim of this paper is to extend the investigation of Ref.~\onlinecite{Mankowsky:2015gv} to evaluate the effect of the third- and fourth-order anharmonic couplings on the structure, electronic and magnetic properties when a phonon is excited by an intense optical pulse. We first use first-principles electronic structure calculations to compute the anharmonic coupling constants between the \unit[17]{THz} polar \PH{B}{1u} phonon mode of \YBCO~with all other phonon modes of the material. We then solve the equations of motion using the first-principles coupling constants, to calculate the structural modulations induced by excitation with an optical pulse. We focus particularly on the case of large amplitude oscillations of the \PH{B}{1u} mode, for which we predict additional structural modulations due to the quartic coupling. Finally, we discuss the requirements for entering the regime in which such interactions dominate, offering a motivation for the provision of THz sources at free-electron lasers.

\section{ANHARMONIC PHONON INTERACTIONS IN YTTRIUM BARIUM COPPER OXIDE}

We begin by using density functional theory to calculate the third- and fourth-order phonon-phonon coupling constants for \YBCO. We choose \YBCO\ since it captures all the characteristics, such as the dopant oxygen atom and the $Pmmm$ orthorhombic symmetry, of the \YBCOsixfive~family, and is computationally convenient because of its small total number of atoms per unit cell. Note that the values for the coupling constants that we obtain differ quantitatively but not qualitatively from those of Ref.~\onlinecite{Mankowsky:2015gv} in which YBa$_2$Cu$_3$O$_{6.5}$ was studied.

First, we calculate the lowest energy atomic positions using the experimental lattice constants of \YBCO~ taken from Ref.~\onlinecite{Andersson:1993uj} ($a=$\unit[3.82]{\AA}, $b=$\unit[3.88]{\AA} and $c=$\unit[11.67]{\AA}). We use the local density approximation (LDA) to density functional theory as implemented within the Vienna \textit{ab-initio} simulation package (VASP) \cite{Kresse:1996vf}, with the default projector augmented wave (PAW) pseudopotentials \cite{Blochl:1994uk} with the following valence electronic configurations: Y ($4s^24p^65s^24d^1$), Ba ($5s^25p^64s^2$), Cu ($3p^64s^13d^{10}$) and O ($2s^22p^4$). After testing the convergence of forces, phonon frequencies and anharmonic coupling constants, we chose a 15$\times$15$\times$10 $k$-point mesh in combination with a cutoff energy of \unit[800]{eV}. Since the ions should be in their equilibrium positions for calculation of the phonons, we relax the internal coordinates until the forces on the ions are less than \unit[0.1]{meV/\AA}. The resulting atomic positions are compared with experiment in Tab.~\ref{tab_struct} and show satisfactory agreement. 

\begin{table}[tb]
\caption{\label{tab_struct} Experimental (EXP) \cite{Andersson:1993uj} and calculated in this work (DFT) atomic positions for \YBCO.}
\begin{ruledtabular}
\begin{tabular}{|l|c|c|c|c|c|}
atom& Wykoff   & x& y	& z (DFT) 	& z (EXP) 		\rule[-1ex]{0pt}{3.5ex} \\
	&  position & 	&	& 			&				\rule[-1ex]{0pt}{3.5ex}\\\hline
Y		& 1h & 	0.5	&0.5		&0.500	& 0.500 \rule[-1ex]{0pt}{3.5ex}\\
Ba		& 2t &	0.5	&0.5		&0.181	& 0.185	\rule[-1ex]{0pt}{3.5ex}\\
Cu$_1$	& 1a&	0.0	&0.0		&0.000	& 0.000	\rule[-1ex]{0pt}{3.5ex}\\
Cu$_2$	& 2q&	0.0	&0.0		&0.353	& 0.355	\rule[-1ex]{0pt}{3.5ex}\\
O$_1$	& 2q&	0.0	&0.0		&0.161	& 0.158	\rule[-1ex]{0pt}{3.5ex}\\
O$_2$	& 2r &	0.0	&0.5		&0.379	& 0.378	\rule[-1ex]{0pt}{3.5ex}\\
O$_3$	& 2s&	0.5	&0.0		&0.379	& 0.378	\rule[-1ex]{0pt}{3.5ex}\\
O$_4$	& 1e&	0.0	&0.5		&0.000	& 0.000	\rule[-1ex]{0pt}{3.5ex}\\
\end{tabular}
\end{ruledtabular}
\end{table}

Next, we calculate the phonon eigenfrequencies and eigenvectors using density functional perturbation theory (DFPT) \cite{BARONI:2001tn}. IR radiation only excites phonon modes at the zone center ($q=0$) and consequently we focus on these modes within our investigation. The \YBCO~unit cell contains 13 atoms and so has 39 phonon modes with the following irreducible representations in the $mmm$ point group: $5A_g \otimes 5 B_{2g} \otimes 5 B_{3g}\otimes 8 B_{1u}\otimes 8 B_{2u} \otimes 8 B_{3u}$. Our calculated mode frequencies are listed in Tab.~\ref{tab_freq_modes}, together with those obtained in other theoretical works \cite{Bohnen:2003ku} as well as from Raman/neutron scattering and IR spectroscopy \cite{Liu:1988es,Burns:1988ti,McCarty:1990uk,Strach:1995hq,Bernhard:2002hf}. We note that the largest difference between our results and other calculations or experiments is less than $\Delta f=$\unit[0.9]{Thz}. 

\begin{table}
\begin{ruledtabular}
\caption{\label{tab_freq_modes}\YBCO~phonon mode frequencies (in THz) obtained in this work (DFT$^*$), calculated using DFT$^\dagger$ in Ref.~\onlinecite{Bohnen:2003ku}, and measured experimentally (EXP) collected from Refs.~\onlinecite{Liu:1988es,Burns:1988ti,McCarty:1990uk,Strach:1995hq,Bernhard:2002hf}.}
\begin{tabular}{|l|r|r|r||l|r|r|r|}
sym.&DFT$^*$&DFT$^\dagger$&EXP&sym.&DFT$^*$&DFT$^\dagger$&EXP\rule[-1ex]{0pt}{3.5ex}\\\hline
\PH{A}{g}	&	3.7	&	3.6	&	3.5	&	\PH{B}{1u}	&	3.9	&	3.9	&	--	\rule[-1ex]{0pt}{3.5ex}\\
	&	4.5	&	4.5	&	4.4	&		&	4.9	&	5.0	&	4.7	\rule[-1ex]{0pt}{3.5ex}\\
	&	10.7	&	10.2	&	10.0&		&	5.9	&	5.8	&	5.8	\rule[-1ex]{0pt}{3.5ex}\\
	&	12.5	&	12.2	&	13.1&		&	7.4	&	6.5	&	--	\rule[-1ex]{0pt}{3.5ex}\\
	&	14.4	&	14.2	&	14.9&		&	7.6	&	7.2	&	--	\rule[-1ex]{0pt}{3.5ex}\\
	&		&		&		&		&	10.0	&	9.7	&	9.0	\rule[-1ex]{0pt}{3.5ex}\\
	&		&		&		&		&	16.9	&	16.5	&	16.9	\rule[-1ex]{0pt}{3.5ex}\\\hline\hline
\PH{B}{2g}	&	2.1	&	1.9	&	2.1	&	\PH{B}{2u}	&	2.7	&	2.5	&	2.5	\rule[-1ex]{0pt}{3.5ex}\\
	&	4.5	&	4.2	&	4.3	&		&	4.5	&	4.0	&	3.8	\rule[-1ex]{0pt}{3.5ex}\\
	&	7.3	&	6.7	&	6.3	&		&	5.0	&	4.6	&	4.7	\rule[-1ex]{0pt}{3.5ex}\\
	&	11.7	&	11.6	&	11.1&		&	5.6	&	5.3	&	5.8	\rule[-1ex]{0pt}{3.5ex}\\
	&	17.9	&	17.3	&	17.4&		&	8.7	&	8.1	&	8.5	\rule[-1ex]{0pt}{3.5ex}\\
	&		&		&		&		&	10.9	&	10.8	&	10.6	\rule[-1ex]{0pt}{3.5ex}\\
	&		&		&		&		&	18.0	&	17.4	&	18.0	\rule[-1ex]{0pt}{3.5ex}\\\hline\hline
\PH{B}{3g}	&	2.4	&	2.4	&	2.5	&	\PH{B}{3u}	&	2.4	&	2.4	&	2.4	\rule[-1ex]{0pt}{3.5ex}\\
	&	4.3	&	4.2	&	4.2	&		&	4.9	&	4.8	&	4.7	\rule[-1ex]{0pt}{3.5ex}\\
	&	9.4	&	8.8	&	9.1	&		&	5.4	&	5.1	&	5.7	\rule[-1ex]{0pt}{3.5ex}\\
	&	11.2	&	11.0	&	11.3&		&	10.5	&	10.1	&	10.3	\rule[-1ex]{0pt}{3.5ex}\\
	&	16.4	&	15.8	&	15.8&		&	10.7	&	10.3	&	10.8	\rule[-1ex]{0pt}{3.5ex}\\
	&		&		&		&		&	15.4	&	14.8	&	14.4	\rule[-1ex]{0pt}{3.5ex}\\
	&		&		&		&		&	16.5	&	15.9	&	16.4	\rule[-1ex]{0pt}{3.5ex}\\
\end{tabular}
\end{ruledtabular}
\end{table}

Finally, we calculate the anharmonic coupling constants. The effective potential describing a polar IR mode (which will be resonantly excited by the pump pulse) and a general phonon mode, G, is to fourth order 
\begin{eqnarray}\label{eq_potentials}
V(\AMP{IR}{},\AMP{G}{})&=&\frac{w^2_{\textnormal{IR}}}{2}\AMP{IR}{}^2+\frac{w^2_{\textnormal{G}}}{2}\AMP{G}{}^2+\\
& &+a_3 \AMP{IR}{}^2\AMP{G}{}+a_4 \AMP{IR}{}^2\AMP{G}{}^2+\nonumber\\
& &+\frac{g^2_{\textnormal{IR}}}{4}\AMP{IR}{}^4+\frac{g^2_{\textnormal{G}}}{4}\AMP{G}{}^4\nonumber \quad,
\end{eqnarray} 
where $w_{\textnormal{IR}}$ and $w_{\textnormal{G}}$ are the frequencies of the IR and G modes respectively, and $a_3$, $a_4$, $g_{\textnormal{IR}}$ and $g_{\textnormal{G}}$ are anharmonic coupling constants (the fourth-order single-mode terms are always real and positive and so by convention their constants are written as $g^2$). As discussed above, the symmetry of the driven mode determines which of the anharmonic coupling constants are non-zero. For the case in which $\AMP{IR}{}$ is a polar mode from the \PH{B}{1u} irreducible representation of the $mmm$ point group, as in the experiments of Ref.~\onlinecite{Mankowsky:2015gv}, only a mode G of \PH{A}{g} symmetry allows $a_3\neq0$, whereas G modes of all symmetry have $a_4\neq0$. 

\begin{figure}
\includegraphics[width=1\columnwidth]{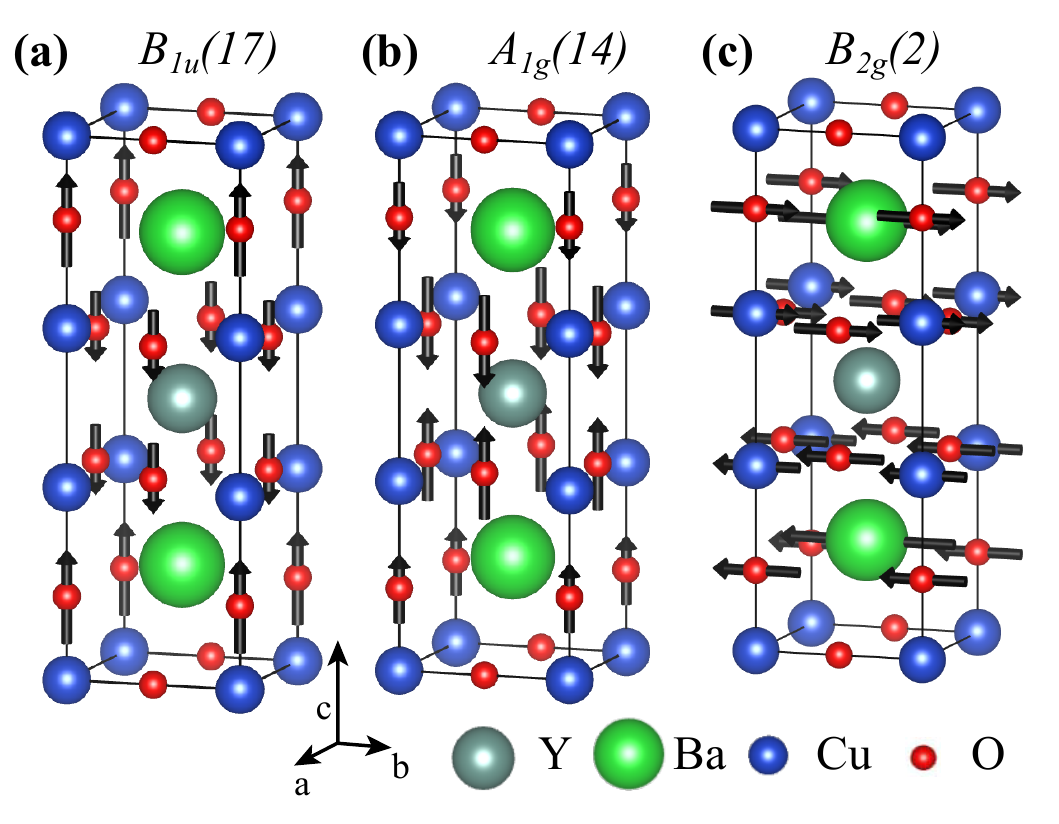}
\caption{\label{fig_modes} Eigendisplacements of selected phonon modes in \YBCO; the black arrows indicate the relative atomic displacements. (a) the pumped polar \PH{B}{1u}(17) phonon mode, (b) the symmetry conserving \PH{A}{g}(14) phonon mode and (c) the in-plane mode \PH{B}{2g}(2).}
\end{figure}

We take the \PH{B}{1u} mode at $f=$\unit[16.9]{THz}, in which the in-plane oxygen atoms displace relative to the Cu atoms along the $c$ axis (Fig.~\ref{fig_modes} (a)) to be the driven IR mode; this is the mode that was driven experimentally in Ref.~\onlinecite{Mankowsky:2015gv}. This mode, which we refer to as \PH{B}{1u}(17) in the following, has non-zero $a_3$ cubic coupling with the five \PH{A}{g} modes and $a_4$ quartic coupling with all 33 modes. We then map the DFT total energies, calculated by freezing in appropriate combinations of phonon modes of known amplitudes, on to the potential expression of Eqn.~\eqref{eq_potentials} to extract the coupling constants by performing least mean square fits of the energy surfaces with the $a_i$ and $g_i$ as the free parameters. 

In Tab.~\ref{tab_couplings} we list the cubic coupling constants between \PH{B}{1u}(17) and all five \PH{A}{g} modes, as well as the quartic couplings for the six $B$ symmetry modes that have a quartic coupling magnitude greater than 0.01 eV/\AA$^3$ with the \PH{B}{1u}(17) mode. To allow for a direct comparison between couplings to modes of different frequencies, and between the cubic and quartic anharmonicities, we make a coordinate transformation $\tilde{Q} = \omega\,Q$ and list also the corresponding renormalized coupling constants $\tilde{a}_3 = \frac{a_3}{\omega_{\textnormal{G}}\omega^2_{\textnormal{IR}}}$  and $\tilde{a}_4 = \frac{a_4}{\omega^2_{\textnormal{G}}\omega^2_{\textnormal{IR}}}$. 

We see that the \PH{A}{g}(14) mode exhibits the largest nominal and renormalized cubic coupling constants and in addition the strongest quartic coupling to \PH{B}{1u}(17). One can understand this qualitatively based on two factors: First its pattern of atomic displacements, shown in Fig.~\ref{fig_modes} (b), is similar to that of the $B_{1u}$(17) mode, with relative Cu-O displacements along the $c$ axis, and second its frequency is closest of all the $A_g$ modes to that of the $B_{1u}$(17) mode. 

\begin{table}[b]
\caption{\label{tab_couplings} Anharmonic coupling constants $a_i$ between $B_{1u}$(17) and other modes. All five $A_g$ modes that show cubic coupling are listed; of the remaining 28 modes only those with a quartic coupling magnitude with $B_{1u}$(17) greater than 0.01 eV/\AA$^3$ are given.}
\begin{ruledtabular}
\begin{tabular}{|c|c|c|c|c|}
\multicolumn{5}{|l|}{cubic}												\rule[-1ex]{0pt}{3.5ex}\\\hline
$f$	& sym. 	& a$_3$& a$_4$		&	$|a_3|$/($\omega_\mathrm{R}\omega^2_\mathrm{IR}$)		\rule[-1ex]{0pt}{3.5ex}\\\hline
[Thz] 	& 	&   [eV/($\sqrt{u}$\AA)$^3$] & [eV/($\sqrt{u}$\AA)$^4$]	&	--			\rule[-1ex]{0pt}{3.5ex}\\\hline
\phantom{1}3.7		&	A$_{g}$		& \phantom{-}0.04 &0.00&0.12	\rule[-1ex]{0pt}{3.5ex}\\ 
\phantom{1}4.5		&	A$_{g}$		& \phantom{-}0.02 &0.01&0.05	\rule[-1ex]{0pt}{3.5ex}\\
12.5				&	A$_{g}$		&			-0.21 &0.04&0.22	\rule[-1ex]{0pt}{3.5ex}\\
14.4				&	A$_{g}$		& \phantom{-}0.70 &0.31&0.61	\rule[-1ex]{0pt}{3.5ex}\\ \hline\hline
\multicolumn{5}{|l|}{quartic}										\rule[-1ex]{0pt}{3.5ex}\\\hline
 $f$				& sym. 			& a$_3$& a$_4$ &$|a_4|$/($\omega^2_\mathrm{R}\omega^2_\mathrm{IR}$)\rule[-1ex]{0pt}{3.5ex}\\\hline
[Thz] 				& 				& [eV/\AA$^2$] & [eV/\AA$^3$]&--\rule[-1ex]{0pt}{3.5ex}\\\hline
\phantom{1}2.1		&	B$_{2g}$	&--& -0.01	&0.17				\rule[-1ex]{0pt}{3.5ex}\\
\phantom{1}4.4		&	B$_{3u}$	&--& -0.01	&0.07				\rule[-1ex]{0pt}{3.5ex}\\
\phantom{1}7.3		&	B$_{2g}$	&--& -0.06	&0.15		   		\rule[-1ex]{0pt}{3.5ex}\\
\phantom{1}8.7		&	B$_{3u}$	&--& -0.07	&0.14		   		\rule[-1ex]{0pt}{3.5ex}\\
\phantom{1}9.4		&	B$_{3g}$	&--& -0.05	&0.08		   		\rule[-1ex]{0pt}{3.5ex}\\
10.7				&	B$_{2u}$	&--& -0.07	&0.10		   		\rule[-1ex]{0pt}{3.5ex}\\ 
\end{tabular}
\end{ruledtabular}
\end{table}

\begin{figure}[t]
\includegraphics[width=1\columnwidth]{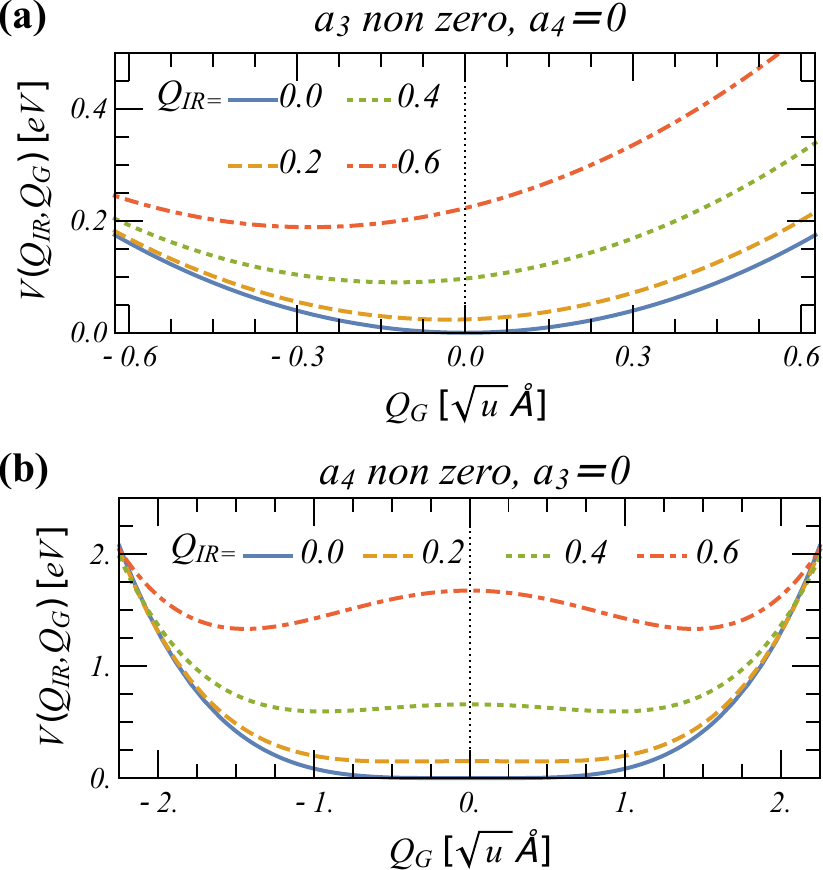}
\caption{\label{fig_potentials} Calculated potential landscapes $V(\AMP{IR}{},\AMP{G}{})$ as a function of the G mode amplitude for different IR mode amplitudes. (a) Cubic coupling of the polar \PH{B}{1u}(17) and the \PH{A}{1g}(14) modes. Excitation of the IR mode displaces the potential minimum for the G mode to a non-zero value. (b) Quartic coupling of the \PH{B}{1u}(17) and \PH{B}{2g}(2) modes. The G-mode potential softens for small IR amplitude, and evolves into a symmetric double well with minima at non-zero G-mode amplitude at large IR amplitude.} 
\end{figure}

To illustrate the changes in the potential landscape caused by the cubic and quartic anharmonicities, we show in Fig.~\ref{fig_potentials} the potential $V(\AMP{IR}{},\AMP{G}{})$ for the \PH{B}{1u}(17) IR mode coupled to (a) the \PH{A}{g}(14) mode and (b) the \PH{B}{2g}(2) mode. In (a) we set the quartic coupling $a_4$ to zero to isolate the effects of the cubic anharmonicity. In both cases the $x$ axis indicates the amplitude of the G mode and the different curves correspond to different amplitudes of the IR mode.
The expected shift of the minimum of the energy well to a non-zero value of G-mode amplitude is seen clearly in the cubic coupling case (upper panel). The sign of the coupling constant determines whether the minimum occurs at positive or negative G-mode amplitude. Since the IR mode amplitudes always appear as squared in the expression for the potential energy, they do not affect the sign of the minimum position, but the magnitude of the minimum shift is larger for larger IR mode amplitude. In the lower panel we see that renormalization of the \PH{B}{2g}(2) mode frequency by the quartic coupling causes it to soften and eventually to become imaginary with increasing IR mode amplitude, indicating a structural instability for large amplitudes.

\section{STRUCTURAL DYNAMICS}

We now turn to the main task of the paper, the determination of the response of the structure following pulsed excitation of the \PH{B}{1u}(17) mode resulting from phonon-phonon coupling. To model this situation we treat each pair of anharmonically coupled phonon modes as coupled classical oscillators \cite{Subedi:2013uh} and solve their equation of motion numerically to obtain the time evolution of both modes after the excitation:
 \begin{equation}\label{eq_eqmotion}
\ddot{\vec{Q}}+ \nabla_{\vec{Q}} \left(V(\AMP{IR}{},\AMP{G}{})-F(t) \AMP{IR}{}\right)=0 \quad.
\end{equation}
Here the vector $\vec{Q}$ contains the eigenvectors of both phonon modes, the dots denote time derivatives and $\nabla_{\vec{Q}}$ is the gradient operator acting on the mode amplitude ($\nabla_{\vec{Q}}$=($\partial/\partial Q_\mathrm{IR},\partial/\partial Q_\mathrm{G}$)). To model the excitation of the IR mode in the manner of the experimental investigations \cite{Mankowsky:2014vt,Mankowsky:2015gv}, we add a pulse-like driving term, $F(t)=F_0 \cos(\omega t) e^{-t^2/(2\sigma^2)}$. Here $F_0$ is the pulse amplitude, $\omega$ the pulse frequency and $\sigma$ the temporal width of the pulse envelope. For ultrashort pulses ($\sigma\rightarrow 0$) the driving term can be described by a $\delta$-function. The experimental pulses, however, exhibit a temporal intensity distribution of \unit[0.3]{ps} (full width at half maximum (FWHM)), which is of the order of the period of the excited polar phonon mode. Consequently, we use a finite $\sigma_0 =$\unit[0.18]{ps} obtained by converting the FWHM value. We set $\omega_{\textnormal{IR}}$ to the frequency of mode \PH{B}{1u}(17) and take the driving force amplitude $F_0=$ \unit[30]{meV/\AA}, which corresponds to a peak electric field strength of \unit[3.0]{MV/cm}, consistent with the experimental value \cite{Mankowsky:2014vt}.

For the cubic coupling case we then solve explicitly the following differential equations:
\begin{eqnarray}\label{eq_cubic1}
	\ddot{Q}_\mathrm{IR}+\freq{IR}{}^2\AMP{IR}{}&=&F(t)-2a_3\AMP{G}{}\AMP{IR}{}-g^2_{\mathrm{IR}}\AMP{IR}{}^3 \\ \label{eq_cubic2}
	\ddot{Q}_\mathrm{G}+\freq{G}{}^2\AMP{G}{}&=&-2a_3\AMP{IR}{}^2-g^2_{\mathrm{G}}\AMP{G}{}^3 \quad,
\end{eqnarray}
neglecting the quartic term, since its inclusion does not alter qualitatively or quantitatively the dynamics. To study the effect of modes that couple only with quartic anharmonicity we solve:
\begin{eqnarray}\label{eq_quartic1}
	\ddot{Q}_\mathrm{IR}+\freq{IR}{}^2\AMP{IR}{}&=&F(t)-2a_4\AMP{G}{}^2\AMP{IR}{}-g^2_{\mathrm{IR}}\AMP{IR}{}^3 \\\label{eq_quartic2}
	\ddot{Q}_\mathrm{G}+\freq{G}{}^2\AMP{G}{}&=&-2a_4\AMP{G}{}\AMP{IR}{}^2-g^2_{\mathrm{G}}\AMP{G}{}^3 \quad. 
\end{eqnarray}

\subsection{Cubic anharmonicities}
\begin{figure*}[t]
\includegraphics[width=2\columnwidth]{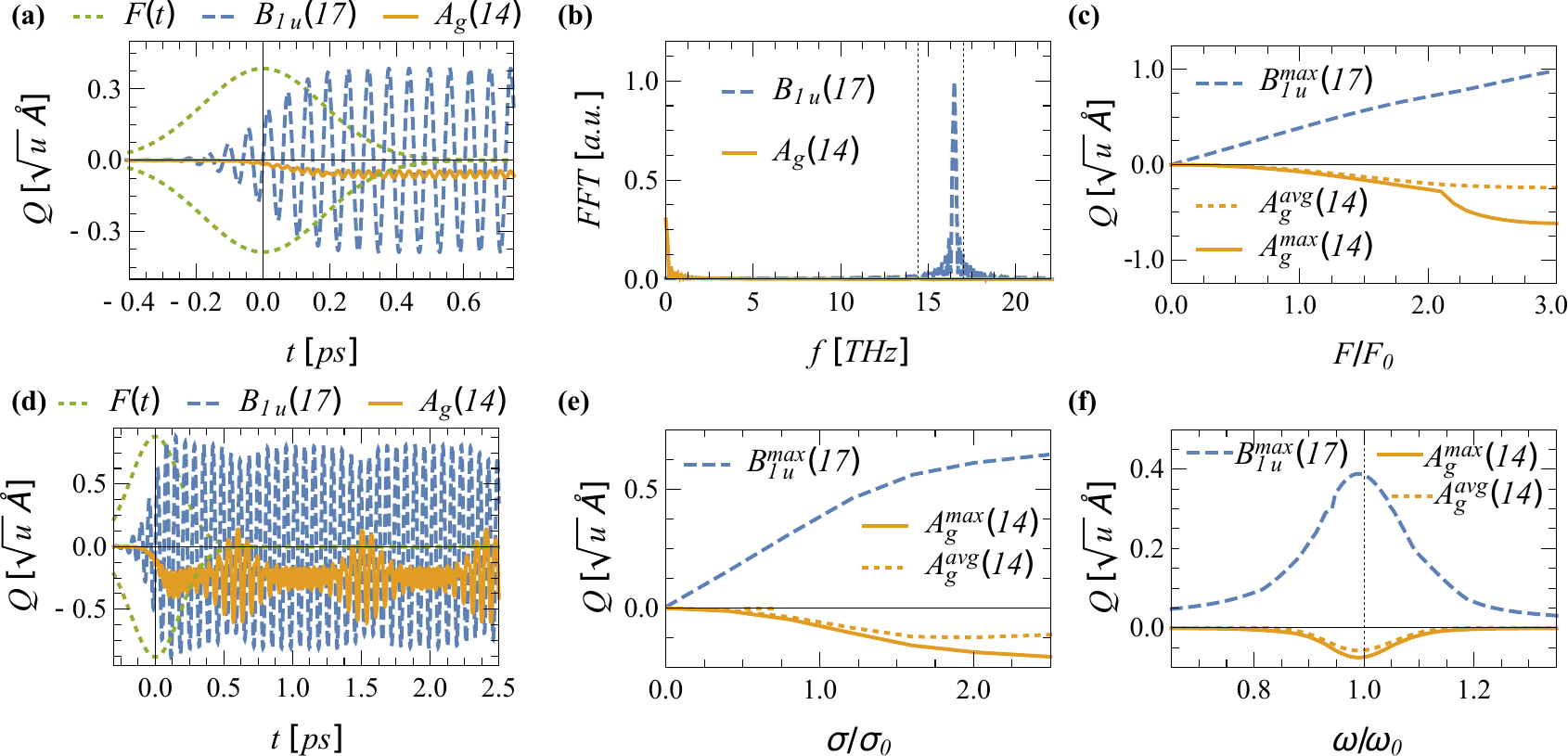}
\caption{\label{fig_cubic} (a) Time evolution of the \PH{B}{1u}(17) (blue dashed line) and  \PH{A}{g}(14) (solid orange line) phonon modes after excitation of the \PH{B}{1u}(17) mode by a pulse $F(t)$ (pulse envelope shown in dotted green). (b) Fourier transform of the time-dependent amplitudes  from (a). The straight lines mark the eigenmode frequencies of the uncoupled \PH{A}{g}(14) and \PH{B}{1u}(17) modes. (c) Maximal amplitudes of the \PH{B}{1u}(17) and \PH{A}{g}(14) modes, and average amplitude of the \PH{A}{g}(14) mode as a function of pump strength, $F$, relative to the reference $F_R$. (d) Time evolution of the \PH{B}{1u}(17) (blue dashed line) and  \PH{A}{g}(14) (solid orange line) after excitation with a pulse of strength \unit[2.5]{$F_0$}. (e) Maximal and average amplitudes as a function of pulse width $\sigma$ relative to the value used in the previous calculation, $\sigma_0=$\unit[0.18]{ps}. (f) Maximal and average amplitudes as a function of pulse frequency, $\omega$ relative to the resonance frequency of the \PH{B}{1u}(17) mode $\omega_0$=\unit[17]{THz}. }
\end{figure*}

We begin by solving the cubic coupling equations (Eqns.~\eqref{eq_cubic1} and \eqref{eq_cubic2}) for the case of the \PH{B}{1u}(14) IR pulse coupled to the \PH{A}{g}(14) G mode. In Fig.~\ref{fig_cubic} (a) we show the time evolution of the amplitudes of the IR and G modes together with the envelope of the pump pulse $F(t)$ (time $t=0$ is set to the maximum  of the pump pulse). We see that on excitation of the \PH{B}{1u} mode, oscillations of the \PH{A}{g}(14) mode are induced through the cubic anharmonic coupling. As shown in Ref.~\onlinecite{Mankowsky:2015gv}, the form of the cubic coupling causes the \PH{A}{g}(14) mode to oscillate around a non-zero displacement, in this case with pulse strength $F_0 = $ \unit[30]{meV/\AA}, the average displacement $Q=$ \unit[-0.03]{$\sqrt{u}$\AA}. This change in average structure has been referred to as a transient structural distortion in the literature.

In Fig.~\ref{fig_cubic} (b) we show the Fourier transforms of the time-dependent amplitudes of both modes. We obtain two main peaks, at frequencies of \unit[16.5]{THz} and \unit[0]{THz} for the \PH{B}{1u}(17) and \PH{A}{g}(14) modes respectively. We see that the frequency of the \PH{B}{1u}(17) mode is shifted by \unit[0.5]{THz} from its eigenfrequency as a result of the anharmonic coupling which renormalizes the frequency according to Eqn.~\eqref{eq_cubic1}:
\begin{equation}\label{eq_mode_shift_cubic}
\tilde{\omega}_{\textnormal{IR}}=\omega_{\textnormal{IR}}\sqrt{1+\frac{2a_3\AMP{G}{}}{\omega_{\textnormal{IR}}}} \quad.
\end{equation}
Since $a_3$ is negative, the frequency decreases as expected. The small zero frequency peak obtained for the \PH{A}{g}(14) mode indicates the static displacement. We note that the small oscillating part of \PH{A}{g}(14) also gives rise to an even smaller peak at 14 THz, however it is much weaker than the zero frequency peak and is not visible on the scale of Fig.~\ref{fig_cubic} (b). 

Next we vary the strength $F_0$ of the pump pulse and show in Fig.~\ref{fig_cubic} (c) the resulting maximum amplitudes of the \PH{B}{1u}(17) (blue line) and \PH{A}{g}(14) (orange line) modes together with the average displacement of the \PH{A}{g}(14) mode (the average displacement of the \PH{B}{1u}(17) mode is always zero). For pump strengths up to the range studied previously, the maximum and average displacements of the \PH{A}{g} mode are equal to each other, consistent with a static off-centering, and follow the linear increase of the maximum amplitude of the \PH{B}{1u} mode. At pump strengths larger than \unit[2.0]{$F_0$}, however, we observe a new behavior, with the maximum amplitude of the \PH{A}{g}(14) mode increasing nonlinearly while its average amplitude starts to saturate. Next we analyze the dynamics of this strong-field behavior. 

The origin of the new behavior of the \PH{A}{g}(14) mode at high pump strength is the shift in frequency of the \PH{B}{1u}(17) pump mode due to the mutual anharmonic coupling. At the pump strength corresponding to the divergence between the maximum and average values of the \PH{A}{g}(14) mode, the frequency of the \PH{B}{1u}(17) mode shifts so that it matches the eigenfrequency of the \PH{A}{g}(14) mode, and the resonant coupling drives the amplified oscillations of the \PH{A}{g}(14) mode. We illustrate this behavior in Fig.~\ref{fig_cubic} (d) which shows the amplitudes of the two modes for $F=$\unit[2.5]{$F_0$}. Coupling between the \PH{A}{g} and \PH{B}{1u} modes causes strong oscillation of the \PH{A}{g} mode amplitude which in turn shifts the frequency of the \PH{B}{1u} mode off resonance so that the \PH{A}{g} amplitude oscillations reduce and the cycle repeats. 

In the remaining two panels of Fig.~\ref{fig_cubic} we show the effect of changing (e) the pulse width, $\sigma$, and (f) the pulse frequency, $\omega$. As expected, we find that the effect of increasing the pulse width is similar to that of increasing the pulse amplitude, since both contribute to an increase in the pulse intensity. Typical resonance behavior for a driven oscillator is seen in the dependence of the responses to the pulse frequency (Fig.~\ref{fig_cubic} (f)) with the amplitudes dropping off rapidly as the pump frequency is moved off resonance (at $\omega = \omega_0$) from the \PH{B}{1u}(17) mode. We see also that the maximum induced amplitude of the \PH{B}{1u}(17) mode occurs at a frequency slightly lower than its eigenfrequency because of its anharmonic coupling to the \PH{A}{g} mode. 

We find that the other phonon modes of \PH{A}{g} symmetry that couple in cubic order to \PH{B}{1u}(17) show similar behavior. For the \PH{A}{g}(4) mode the induced quasi static off-centering amplitude is \unit[0.01]{$\sqrt{u}$\AA} after excitation of \PH{B}{1u} by our reference pulse with strength $F_0$. For the two other \PH{A}{g} modes the static off-centered amplitudes are half of this size. The dynamics of the mode oscillation as a function of pulse strength, frequency and width is qualitatively similar for all modes. We note finally that in a real system the excitation of the \PH{B}{1u}(17) mode generates a structure that is a superposition of the displacements caused by all relevant \PH{A}{g} modes, rather than just the one considered here.

Finally, we discuss our results with respect to the experimental findings of Ref.~\onlinecite{Mankowsky:2014vt}. In Ref.~\onlinecite{Mankowsky:2014vt}, time-resolved x-ray diffraction on optically pumped \YBCOsixfive~found a quasi-static structural change corresponding to a reduction in the Cu - apical oxygen distance of \unit[2.2]pm, which was attributed to cubic anharmonic phonon-phonon coupling. This is the same distortion pattern that we obtain in this work for YBa$_2$Cu$_3$O$_7$, suggesting that it should be found across the entire yttrium barium copper oxide series, and is at least qualitatively independent of the doping concentration. We note, however, that our calculations do not capture the second finding of Ref.~\onlinecite{Mankowsky:2014vt}, of induced changes in the inter- and intra-plane distances, since the smaller unit-cell size that we use in this work does not allow this degree of freedom. 

\subsubsection{Induced changes in electronic and magnetic properties}

We now discuss the effect of the structural modulations induced by phonon excitation on the electronic and magnetic properties. We focus in particular on the changes in the magnetic exchange interactions and the density of states at the Fermi level, both of which are believed to be relevant for superconductivity in cuprate superconductors \cite{Pickett:1989tk}. We distinguish between the effect of the change in the time-averaged structure associated with the shift in the minimum of the potential well to a non-zero value of the \PH{A}{g}(14) mode previously referred to as a transient structural distortion (we call this the quasi-static contribution), and the ongoing oscillations around this average structure (which we call the oscillatory contribution) that are dominated by the \PH{B}{1u}(17) mode. 

\begin{figure}[tb]
\includegraphics[width=1\columnwidth]{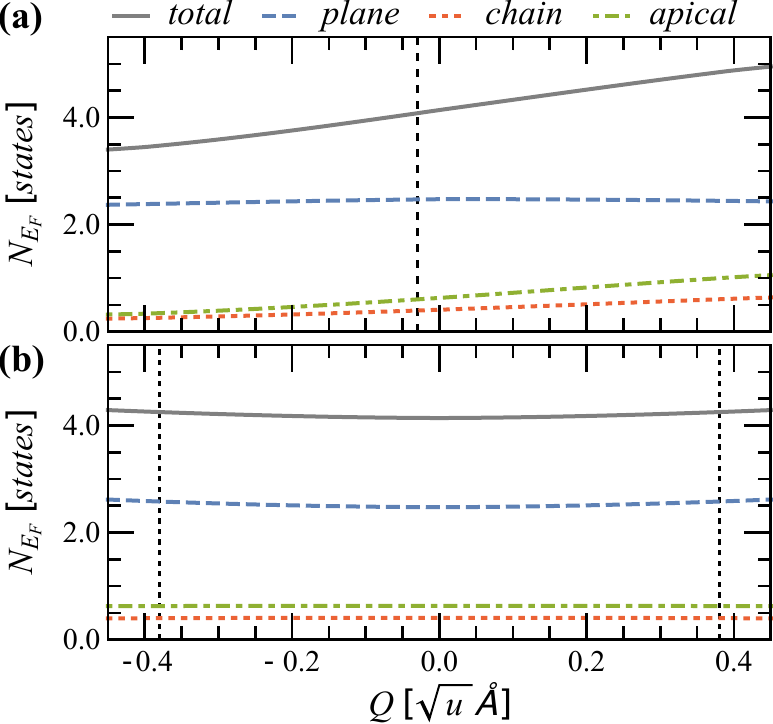}
\caption{\label{fig_dos} Calculated densities of states at the Fermi level, $N_{E_F}$, as a function of (a) the \PH{A}{g}(14) mode amplitude and (b) the \PH{B}{1u}(17) mode amplitude. The dashed vertical lines indicate the values of the averaged \PH{A}{g}(14) mode amplitude (in (a)) and \PH{B}{1u}(17) maximum amplitude (in (b)) at our reference pump strength.} 
\end{figure}

We begin with the density of states at the Fermi energy, $N_{E_F}$.  In the equilibrium structure, $N_{EF}$=\unit[4.14]{states}, with the majority contribution (\unit[2.47]{states}) coming from the copper-oxygen planes and smaller contributions from the copper ions in the chains (\unit[0.64]{states}) and the apical oxygens (\unit[0.41]{states}). (Note that the difference between the total and local values occurs because the local contributions are obtained from projecting the density of states into atomic spheres, so interstitial contributions are not captured.) Fig.~\ref{fig_dos} (a) shows the change in $N_{E_F}$ as a function of the amplitude of the \PH{A}{g}(14) mode, with the vertical line indicating the value of the quasi-static distortion at our reference pump strength. We find a reduction in electron count of around \unit[-1.4]{\%} for the negative amplitudes that are present in the quasi-static structure; the density of states in the copper-oxygen planes is largely unchanged however. In Fig.~\ref{fig_dos} (b) we show the corresponding variation in $N_{E_F}$ as a function of the \PH{B}{1u}(17) mode amplitude, with the vertical lines indicating the maximum amplitudes at our reference pulse strength. Here, $N_{E_F}$ increases quadratically with \PH{B}{1u}(17) mode amplitude, reaching $\sim$ \unit[2.7]{\%} difference at the maximum amplitude, with the change dominated by the states in the CuO$_2$ planes. Consequently, although the time averaged amplitude of the polar \PH{B}{1u}(17) mode is zero, its oscillation causes a change in $N_{E_F}$ comparable in magnitude to that caused by the quasi-static distortion and likely more relevant for the description of the physics in the copper-oxygen planes.

\begin{figure}[tb]
\includegraphics[width=1\columnwidth]{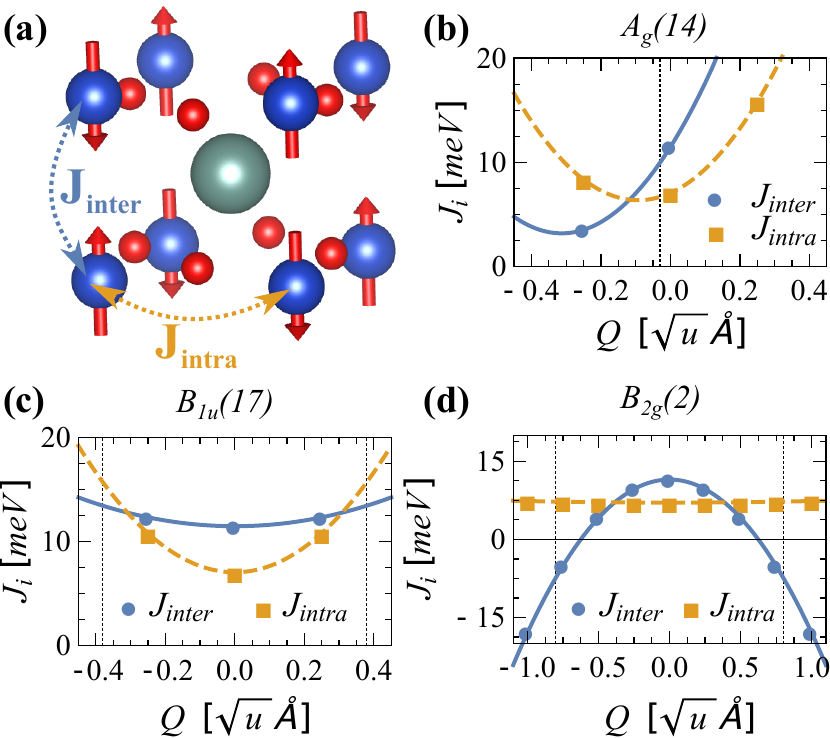}
\caption{\label{fig_exchanges} (a) Schematic of the two copper-oxygen planes of \YBCO\ indicating the two magnetic exchanges, $J_{inter}$ and $J_{intra}$, that we consider in this work.  (b) - (d) Calculated magnetic exchange interaction as a function of mode amplitude for (b) \PH{A}{g}(14) (c) \PH{B}{1u}(17) and (d) \PH{B}{2g}(2) mode amplitudes. The vertical dashed lines in (b) and (c) indicate the values of average and maximum mode amplitude induced by the reference pulse for \PH{A}{g}(14) and \PH{B}{1u}(17), respectively. In (d) the vertical lines show the induced maximum amplitude of \PH{B}{2g}(2) after excitation of the system with a field pulse four times stronger than the reference pulse. Note the different $y$-axis scales.}
\end{figure}

Next, we evaluate the magnetic exchange interactions by mapping our calculated DFT total energy differences onto a Heisenberg model using the approach of Refs.~\onlinecite{Xiang:2011cn,Wagner:2014js}. We use a simple Heisenberg model with only two magnetic exchanges, $J_{inter}$ and $J_{intra}$ (Fig.~\ref{fig_exchanges} (a)), as measures of the exchange between and within the copper-oxygen planes, respectively. In the ground-state structure we find that both exchanges are antiferromagnetic with $J_{inter}$=\unit[11.5]{meV} and $J_{intra}=$\unit[7.0]{meV}. (Note that these values for the overdoped system are smaller than those for optimally doped cuprates, which are typically in the order of \unit[100]{meV}.) In Fig.~\ref{fig_exchanges} (b) we show $J_{inter}$ and $J_{intra}$ as a function of the \PH{A}{g}(14) mode amplitude, again with the vertical dashed line indicating the amplitude at the quasi-static structure induced by our reference pulse. Both $J_{inter}$ and $J_{intra}$ decrease, by around \unit[13]{\%} and \unit[5]{\%} respectively. In Fig.~\ref{fig_exchanges} (c) we show our calculated values of $J_{inter}$ and $J_{intra}$ as a function of the polar \PH{B}{1u}(17) mode amplitude, again with the vertical lines indicating the maximum amplitudes at our reference pulse strength. We find a quadratic increase of both magnetic exchanges with mode amplitude, with $J_{inter}$ and $J_{intra}$ increasing by \unit[17]{\%} and \unit[122]{\%} respectively at the maximum amplitude of the \PH{B}{1u}(17). These oscillatory changes exceed those found for the quasi-static structure.

To summarize this section, we find that excitation of the \PH{B}{1u}(17) phonon mode causes a quasi-static decrease in both the magnetic exchanges and the density of states at the Fermi energy, accompanied by strong oscillations of these values.

\subsection{Quartic anharmonicities}
\begin{figure*}[tb]
\includegraphics[width=2\columnwidth]{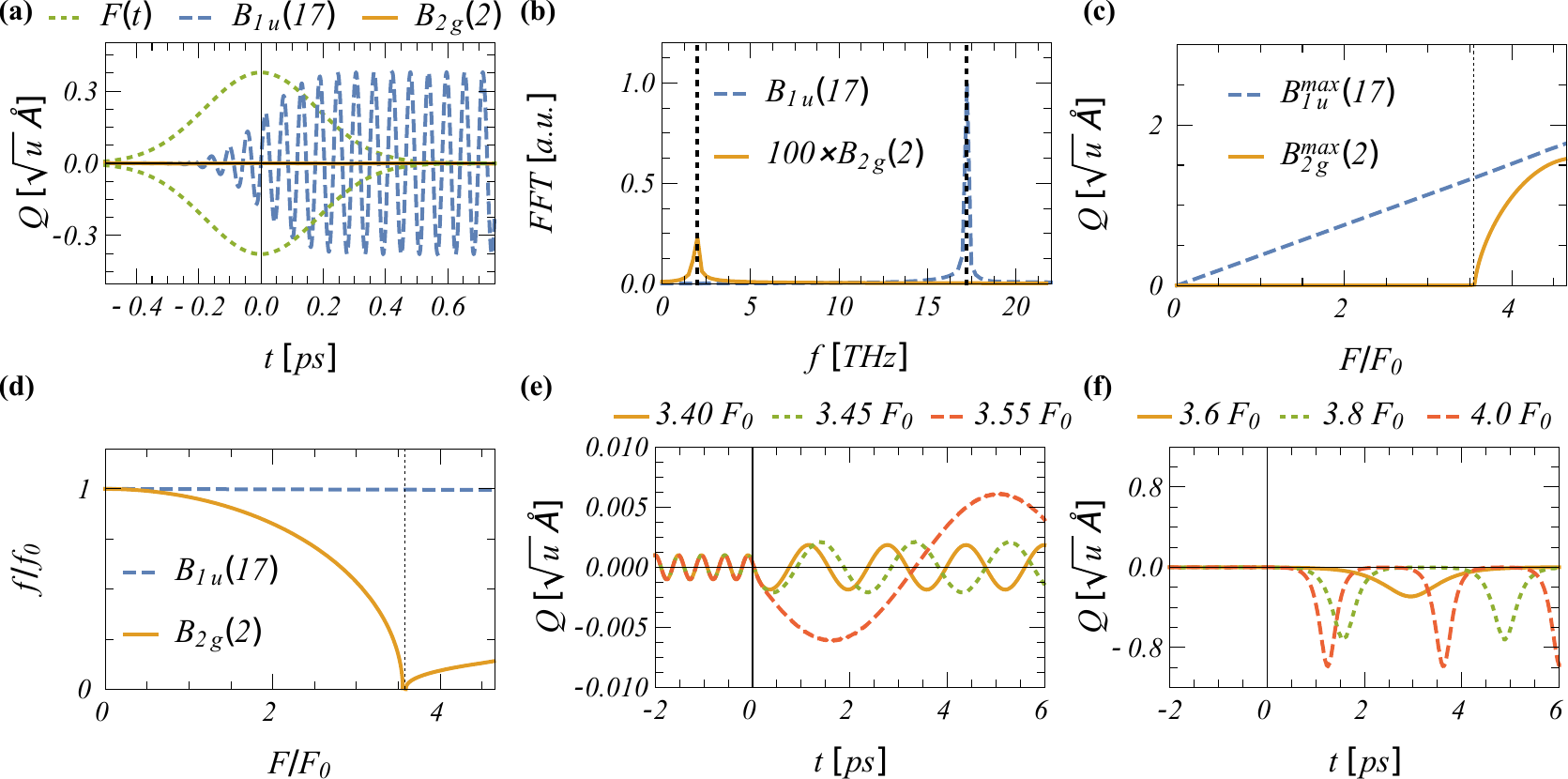}
\caption{\label{fig_quartic} (a) Time evolution of the \PH{B}{1u}(17) (blue dashed) and \PH{B}{2g}(2) (orange) phonon modes after excitation of the \PH{B}{1u}(17) mode by a pulse $F(t)$ (pulse envelope shown in dotted green). (b) Fourier transform of the time-dependent amplitudes in (a). The straight lines mark the mode frequencies of \PH{B}{2g}(2) and \PH{B}{1u}(17). Note that the spectra of \PH{B}{2g}(2) is scaled by a factor of 100. (c) Maximum amplitudes of the \PH{B}{2g}(2) and \PH{B}{1u}(17) modes after excitation with different pulse strengths $F$ given relative to the reference value $F_0=$\unit[3]{MV/cm}. (d) Oscillation frequencies $f$ of the \PH{B}{2g}(2) and \PH{B}{1u}(17) modes (relative to their unperturbed eigenfrequencies $f_0$) as a function of excitation pulse strength. (e,f) Time evolution of the \PH{B}{2g}(2) mode for a range of \PH{B}{1u}(17) mode pulse strengths below (e) and above (f) the critical strength at which where the frequency of the \PH{B}{2g}(2) mode becomes zero. Note that the time range is extended with respect to that shown in (a).}
\end{figure*}

In this section we analyze the coupling between the pumped \PH{B}{1u}(17) mode and the coupled $B$ modes that do not allow cubic coupling by symmetry in order to isolate the effects of the quartic anharmonicities. Since quartic coupling renormalizes the mode frequencies according to Eqn.~\eqref{eq_mode_renormalization} we focus on the \PH{B}{2g}(2) mode which has the largest renormalization due to its low frequency. In Fig.~\ref{fig_quartic} (a) and (b) we show respectively the time dependence of the amplitudes and the corresponding Fourier transforms for the \PH{B}{2g}(2) and \PH{B}{1u}(17) modes after excitation of the polar mode by our reference pulse of strength $F_0 =$ \unit[3]{MV/cm}. At this pulse strength, we see that the quartic anharmonic coupling induces no noticeable oscillation of the \PH{B}{2g}(2) mode. The Fourier transform of the oscillation pattern shows a peak at the original \PH{B}{1u}(17) mode eigenfrequency and a tiny peak at the \PH{B}{2g}(2) frequency  which is hardly visible on this scale. Both frequencies are indicated by the vertical black lines in Fig.~\ref{fig_quartic} (b). 

Next we investigate the effect of the pump strength and show in Fig.~\ref{fig_quartic} (c) the evolution of the maximum amplitudes for both modes as a function of the pump strength, $F$ relative to the reference value $F_0$. (Note that, since the cubic coupling is zero by symmetry, the average displacement of both modes is always zero.)  As expected, the maximum amplitude of the \PH{B}{1u}(17) mode increases linearly with the pulse strength. The behavior of the \PH{B}{2g}(2) mode, however, is strikingly different. Its maximum amplitude is at first independent of pump strength (on this scale it is indistinguishable from zero), until at $F=3.6 F_0=$\unit[10.8]{MV/cm} it shows an abrupt nonlinear increase, rapidly becoming comparable to the amplitude of the \PH{B}{1u}(17) mode. To understand this behavior, we show in Fig.~\ref{fig_quartic} (d) the frequencies of the \PH{B}{1u}(17) and \PH{B}{2g}(2) modes as a function of normalized pulse strength. We see that increasing the pulse strength does not affect the frequency of \PH{B}{1u}(17) over the range studied. For \PH{B}{2g}(2), however, we find that the frequency {\it decreases} with pulse strength, corresponding to the renormalization due to the quartic coupling (Eq.~\eqref{eq_mode_renormalization}) and reaches zero at the pulse strength ($\sim 3.6\,F_0$) corresponding to the sharp increase in the maximum amplitude. At higher pump strength the \PH{B}{2g}(2) mode frequency is again finite and increases slightly. 

To better understand the nature of the oscillations in the range of the critical pump strength we show in Fig.~\ref{fig_quartic} (e) and (f) the time-dependent amplitudes of the \PH{B}{2g}(2) mode for pulse strengths slightly below and above the critical value of $F=3.6\,F_0$ respectively. For pulse strengths below the critical value we see that the oscillations remain sinusoidal but show a strong increase in wavelength and amplitude as the critical value is approached. In contrast, for pulse strengths above the critical value, the response is non-sinusoidal and the oscillation amplitudes are substantially higher (note the different $y$-axis scale) due to the dynamical instability of the \PH{B}{2g}(2) mode. These large amplitude oscillations of the \PH{B}{2g}(2) mode have a drastic effect on the \YBCO~ structure. From Fig.~\ref{fig_modes} (c), which shows the eigendisplacement of the \PH{B}{2g}(2) mode, we see that the induced structural change is a shear of adjacent CuO$_2$ bilayers in opposite directions in the $a-b$ plane (the two CuO$_2$ layers within a bilayer shear in the same direction). The oscillations, shown in Fig.~\ref{fig_quartic} (f), reach amplitudes of \unit[0.8]{$\sqrt{u}$\AA}, which correspond to movements of the atoms in the BaO plane of about \unit[5]{pm} and of about \unit[2]{pm} relative motion of the BaO and CuO$_2$ planes. Interestingly, the frequency renormalization of the \PH{B}{2g}(2) mode given by Eqn.~\eqref{eq_mode_renormalization} suggests imaginary frequencies for amplitudes larger than $Q^\mathrm{crit}_\mathrm{IR}=$\unit[0.91]{$\sqrt{u}$\AA} of the \PH{B}{1u}(17) mode. In our simulations, however, we find that the instability manifests only at amplitudes of the \PH{B}{1u}(17) above $Q_\mathrm{IR}=$\unit[1.35]{$\sqrt{u}$\AA} (Fig.~\ref{fig_quartic} (c) and (d)). This corresponds to the pulse strength at which the amplitude of the \PH{B}{1u}(17) mode exceeds the critical amplitude for around half of the time; a maximum amplitude equal to the critical value is insufficient.  For further increasing amplitudes of \PH{B}{1u}(17), the time that must be spent above the critical value in order for the instability to manifest increases as $t_\mathrm{crit.}=acos(Q_\mathrm{IR}^\mathrm{crit}/Q_\mathrm{IR})$, and eventually saturates, reflected within Fig.~\ref{fig_quartic} (d) in the range beyond $F=3.6\;F_0$. (This last result is obtained trivially from the time that a sinusoidal oscillation spends above a specific amplitude.)

Finally, we discuss the oscillations in the electronic properties caused by the quartic coupling to the \PH{B}{2g}(2) mode for a pulse of $3.8\;F_0$ (Fig.\ref{fig_quartic} (f)). Since the \PH{B}{2g}(2) mode eigenvector consists of a shear of the two copper-oxygen planes in the bilayer we expect a strong effect on the inter-plane exchange interactions which we indeed see in Fig.~\ref{fig_exchanges} (d): at maximum amplitude $J_{inter}$ reduces by 164 \% to \unit[-7.3]{meV}. $J_{intra}$ is largely unaffected by excitation of the \PH{B}{2g}(2) mode, as is the density of states at the Fermi energy (not shown). We note that this shear motion induces an ultrafast oscillating stress along the $b$ direction. We extract the value of this stress using the frozen phonon method and obtain a value of $\sim$\unit[1]{GPa} for an amplitude of \unit[0.8]{$\sqrt{u}$\AA} of the \PH{B}{2g}(2) mode. In addition to the distortion of the \PH{B}{2g}(2) mode, fluctuations of the exchange values caused by the \PH{B}{1u}(17) mode also occur here, at the correspondingly higher frequency. Consequently, strong pumping will cause both a temporal slow sign change of the inter-layer exchange interactions accompanied with fast oscillations of the intra-layer exchanges. 

\section{Summary and Discussion}

In summary, we have computed the anharmonic phonon-phonon couplings up to fourth order in \YBCO~and explored the structural dynamics induced by pulsed pumping of an IR phonon mode. Consistent with previous work \cite{Mankowsky:2014vt}, we find that cubic phonon-phonon coupling of type $Q^2_\mathrm{IR}Q_\mathrm{G}$, with G a mode of \PH{A}{g} symmetry, dominates at lower excitation strengths. In addition, we find several low-frequency modes that exhibit a sizable quartic coupling of the form $Q^2_\mathrm{IR}Q^2_\mathrm{G}$. Our computations of the effect of pulsed excitation of an IR mode reveal that various kinds of structural dynamics can be triggered in \YBCO, depending on the pulse strength. For low pulse strengths ($F\lessapprox$\unit[3]{MV/cm}), the cubic coupling causes a quasi-static modulation of the structure, whereas the quartic is not significant.
 
For moderate to large pulse strengths ($F\gtrapprox$\unit[6]{MV/cm}), the quasi-static displacement caused by the cubic coupling exhibits additional large-amplitude modulations of the \PH{A}{g} mode. In addition, for even stronger pulse strengths ($F\gtrapprox$\unit[10.5]{MV/cm}) the quartic coupling becomes activated and the coupled \PH{B}{g}(2) mode become imaginary, at which point its oscillation becomes non-sinusoidal with large amplitude. These effects were not discussed in earlier work, which used lower pulse strengths, but are accessible for example with free-electron lasers, for which field strengths reach up to \unit[50]{MV/cm}. Indeed our predicted induction of a high-frequency dynamical shear strain at high field provides a motivation for provision of THz sources at free-electron lasers to allow experimental exploration of this regime, where measurement of the induced changes in the the superconducting behavior could be valuable in understanding the pairing mechanism. 

In particular, we saw that while the induced changes in the quasi-static structure does not have a strong effect on the electronic properties, the oscillating changes, particularly along the eigenvector of the \PH{B}{1u} mode, strongly modify the magnetic exchange interactions, which are likely relevant for superconductivity \cite{Pickett:1989tk}. While we emphasize that our calculations are for the overdoped regime, the strong sinusoidal oscillation that our simulations reveal in the intra- and interplanar magnetic exchange interactions could be relevant for the observed signatures of coherent transport above the equilibrium critical temperature \cite{Kaiser:2012ux,Hu:2013vw}. In addition to the superconducting properties, it was recently suggested that coupled spin-lattice fluctuations could be a source of magnetic quadrupolar order related to the pseudogap phase of cuprate oxides \cite{Fechner:2015va}. Whether such an order is suppressed or enhanced by an additional induced fluctuation of the magnetic exchange interaction is an intriguing question, and its experimental resolution could shed additional light on the relevance of spin-lattice coupling in cuprates. 

Finally, we note that the link between the dynamic and quasistatic structural changes of either cubic or quartic anharmonic origin and the magnetic exchanges suggests non-linear phononics as a route to novel coupled phonon-magnon behavior. An example of such a coupling could be the recently published work of Ref.~\onlinecite{Nova:2015ux}, in which intense optical excitation of two orthogonal phonon modes has been shown to excite a magnon. 

\section{Acknowledgements}

This work was supported by the ETH Zurich, by the ERC Advanced Grant program, No. 291151 and by the NCCR MARVEL, funded by the Swiss National Science Foundation. Calculations were performed at the Swiss National Supercomputing Centre (CSCS) under project ID s624. We thank Andrea Cavalleri, Antoine Georges and Roman Mankowsky for useful discussions. 

\bibliography{ybco_phonon,misc}

\end{document}